\documentclass[aps,prb,reprint,superscriptaddress,showpacs,showkeys]{revtex4-1}

\usepackage{amssymb}
\usepackage{amsmath}
\usepackage{emf}
\usepackage{amsthm}
\usepackage{mathtools}
\usepackage{bbold}
\usepackage{enumitem}
\usepackage{tikz}
\usepackage{circuitikz}

\usetikzlibrary{shapes.geometric,arrows,calc}
\usetikzlibrary{arrows.meta,decorations.markings,plotmarks}
\usepackage{float}

\newcommand{\bs}[1]{\boldsymbol{#1}}
\newcommand{\od}{\mathrm{d}\mkern0mu}

\newtheorem*{remark*}{Remark}

\begin{document}


\title{Concepts of Electric Circuit Analysis Revised: Analysis without Current Direction}


\author{Toma\v{z} Podobnik}
\affiliation{University of Ljubljana, 
Faculty of Mathematics and Physics,
Jadranska 19, 1000 Ljubljana, Slovenia}
\affiliation{Jo\v{z}ef Stefan Institute, Jamova 39, 1000 Ljubljana, Slovenia}


\date{\today}

\begin{abstract}
The electric current as the flux of current density---a signed scalar, not a vector---is inconsistent with 
the concept of the current direction,  commonly invoked in the electric circuit analyses within, 
for example, Kirchhoff's current law, the resistance rule, and Lenz's law. This paper presents an approach 
to the circuit analysis that is not predicated upon the ``current direction,'' and  thus avoids such 
inconsistencies. The approach is summarized in five  rules  and the application of these  rules is demonstrated 
by analyzing an RLC series circuit. 
\end{abstract}
%


\maketitle

%

\section{Introduction: the self-contradictory concept of current direction}
\label{sec:introduction}
The electric current $I$ is the flux (surface integral) of the current density vector field 
$\mathbf{j}(\mathbf{r})$,\!\cite{feynman,grant,purcell,griffiths,walker}
\begin{equation}
\label{eq:gendefI}
I\coloneqq \int\limits_{S}\!\mathbf{j}(\mathbf{r})\cdot \mathop{\od\mathbf{S}},\tag{1.a}
\end{equation}
which, for a uniform $\mathbf{j}(\mathbf{r})$ and a flat surface $S$, simplifies to
\begin{equation}
\label{eq:currexam}
I = \mathbf{j}\cdot \mathbf{S} = {j}A\mathop{\hat{\mathbf{e}}_{\mathbf{j} }\cdot \hat{\mathbf{n}}} . \tag{1.b}
\setcounter{equation}{1}
\end{equation}
Above, $j$ is the magnitude of $\mathbf{j}$ and $\hat{\mathbf{e}}_{\mathbf{j}}=\mathbf{j}/j$ is its direction
(the drifting direction of positive charge carriers), $\mathbf{S}= A\mathop{\hat{\mathbf{n}}}$ is the surface vector, 
$A$ is the area of $S$, and $\hat{\mathbf{n}}$ is its direction (orientation), a unit vector, perpendicular to $S$; there 
are always {\em two} such vectors, opposite to each other, that we must {\em choose} from for $I$ to be specified.

By Definition\;\eqref{eq:gendefI}, $I$ is {\em a signed scalar,} {\em not} a vector,\!\cite{feynman1} and is thus  
incompatible with the concept of current direction, commonly invoked in the electric circuit analyses within, for example,
Kirchhoff's current law,\!\cite{walkerKirchhoff} the resistance rule,\!\cite{walkerresistance} and 
Lenz's law.\!\cite{feynmanLenz,grantLenz,purcellLenz,griffithsLenz} This paper presents an approach to the circuit analysis 
that is {\em not} predicated upon the ``current  direction'' and thus avoids such inconsistencies. The approach is valid within 
the lumped-element approximation,\!\cite{feynmanlumped} according to which the time-varying magnetic field (its flux) in 
electric circuits is negligible everywhere {\em outside} the elements, as well as {\em inside resistors} and {\em capacitors}.
In Sec.\;\ref{sec:rules}, the concepts are summarized in five rules, (R.1)--(R.5), and the application of these rules is demonstrated  
in Sec.\;\ref{sec:application} by analyzing  an RLC series circuit. The derivation of the rules is provided in the Appendix. 
%
%
\section{Five rules}
\label{sec:rules}
%
%
%

Let $V$ be a junction of $m$ wires, with $I_i$ being the electric current through the $i$-th wire 
(through the contact surface of the wire and the junction) (Fig.\,1);  the (cross-sections of the) 
wires (their contact surfaces with $V$) numbered $1$ to $k$ are oriented toward the junction 
interior, and the remaining wires (contact surfaces) are oriented outward.
\begin{figure}[htbp]
\label{fig:junction}
\vskip -2.25cm
\hspace*{-4.75cm}
\includegraphics[width=18cm]{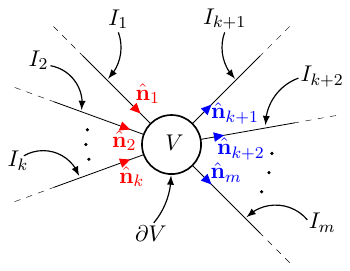}
\vskip -17.7cm
\caption{A junction $V$ of $m$ wires. The (contact surfaces between the junction and the) wires numbered $1$ to $k$ are oriented 
toward the junction interior, and the remaining wires (contact surfaces) are oriented outward; the orientations are set arbitrarily, 
independently of the (possibly unknown) drifting directions of charge carriers through particular surfaces (into the junction or out of it). 
Currents $I_i$ through the wires (through the contact surfaces), on the other hand, are scalars and thus have {\em no} direction.}
\end{figure}
The change rate $\dot{q}^{}_{V}$ of the total charge in $V$ is given by the continuity equation in the integral form, 
\begin{equation}
\label{eq:R1}
\boxed{\dot{q}^{}_{V}  =
\sum_{i=1}^{k}{\!I_i}  -\!\!\sum_{i=k+1}^{m} \!\!I_i\ ;\ 
\hat{\mathbf{n}}_i: 
\begin{cases}
\textnormal{into }V &\hskip -3mm ; i\le k \\
\textnormal{out of }V &\hskip -3mm ; i>k
\end{cases}\, .} \tag{R.1}
\end{equation}
%
%
%

Consider an electric circuit consisting of lumped elements (Fig.\,2), and a loop $\partial S'$ consisting of the paths $\mathcal{C}'_i$
{\em by} ({\em not through}) the elements, $\partial S' = \sum_i \mathcal{C}'_{i}$. 
The voltage (i.e., minus the integral of the electric field $\mathbf{E}$) along the path $\mathcal{C}'_i$ is $U'_i$, and 
\begin{figure}[htbp]
\vskip -2.7cm
\hspace*{-5.85cm}
\includegraphics[width=20cm]{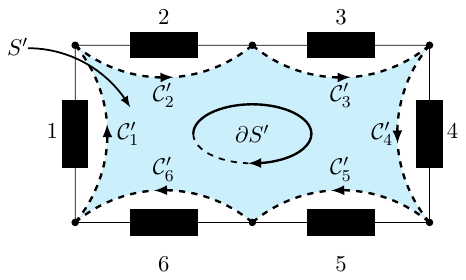}
\vskip -19.35cm
\caption{An electric circuit consisting of several lumped ele\-ments.\!\cite{feynmanfigure} The surface $S'$ lies entirely in the region with 
negligible time-varying magnetic field. The boundary $\partial S'$ of the surface does {\em not} pass through any of the elements.}
\end{figure}
the sum of the voltages around the loop is zero,
\begin{equation}
\label{eq:R2}
\boxed{\sum\limits_i U'_i=0\, ,}\tag{R.2}
\end{equation}
which is Kirchhoff's voltage law (loop rule).\!\cite{feynmanKirchhoff} 

The remaining three rules specify the voltages along the paths
{\em by} resistors, capacitors, and inductors.
%
%
%
The voltage $U'_{\mathrm{R}}$ along a path $\mathcal{C}'_{\mathrm{R}}$ from point $P_1$ to point $P_2$,
passing {\em by} a resistor of resistance $R$ (Fig.\,3.(a)), is
\begin{figure}[htbp]
\vskip -1.95cm
\hspace*{-3.9cm}
\includegraphics[width=16.3cm]{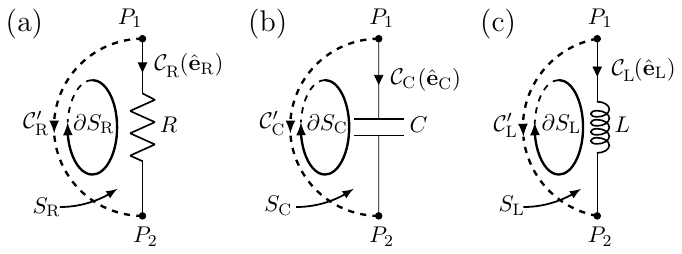}
\vskip -16.15cm
\caption{(a) A loop $\partial S_{\mathrm{R}}$ (the boundary of a surface $S_{\mathrm{R}}$) consisting of the paths $\mathcal{C}_{\mathrm{R}}$ 
and $-\mathcal{C}'_{\mathrm{R}}$  {\em through} and {\em by}  a resistor; $\hat{\mathbf{e}}_{\mathrm{R}}$ is the orientation (direction of a differential 
element) of the path $\mathcal{C}_{\mathrm{R}}$. (b) and (c) Analogously for the loops $\partial S_{\mathrm{R}}$ and $\partial S_{\mathrm{L}}$
comprising a capacitor and an inductor.}
\end{figure}
\begin{equation}
\label{eq:R3}
\boxed{U'_{\mathrm{R}} =-\hat{\mathbf{n}}_{\mathrm{R}}\cdot\hat{\mathbf{e}}_{\mathrm{R}}\mathop{I_{\mathrm{R}} R}\, ,}\tag{R.3}
\end{equation}
where $\hat{\mathbf{e}}_{\mathrm{R}}$ is the orientation (direction of a differential element) of the path $\mathcal{C}_{\mathrm{R}}$ from
$P_1$ to $P_2$ {\em through} the resistor and along its connecting wires, $\hat{\mathbf{n}}_{\mathrm{R}}$ ($+\hat{\mathbf{e}}_{\mathrm{R}}$ 
or $-\hat{\mathbf{e}}_{\mathrm{R}}$) is the chosen orientation of the wires and the resistor (their cross-sections), 
and $I_{\mathrm{R}}$ is the current through the resistor.
%
%
%

The voltage $U'_{\mathrm{C}}$ along a path $\mathcal{C}'_{\mathrm{C}}$ 
by a capacitor of capacitance $C$ (Fig.\,3.(b)) is
\begin{equation}
\label{eq:R4}
\boxed{U'_{\mathrm{C}}=-\frac{q_{\mathrm{C},1}}{C}=+\frac{q_{\mathrm{C},2}}{C}\, ,}\tag{R.4}
\end{equation}
where $q_{\mathrm{C},1}$ is the charge of the electrode, connected to the starting point $P_1$ of $\mathcal{C}'_{\mathrm{C}}$, and $q_{\mathrm{C},2}\;(=-q_{\mathrm{C},1})$ 
is the charge of the electrode, connected to the endpoint $P_2$.
%
%

Finally, the voltage $U'_{\mathrm{L}}$ along a path $\mathcal{C}'_{\mathrm{L}}$ {\em by} an inductor of inductance $L$ (Fig.\,3.(c)) is
\begin{equation}
\label{eq:R5}
\boxed{ U'_{\textnormal{L}}=
-\mathop{\hat{\mathbf{n}}_{\textnormal{L}} \cdot\hat{\mathbf{e}}_{\textnormal{L}}}L\dot{I}_{\textnormal{L}}\, ,} \tag{R.5}
\end{equation}
where $\hat{\mathbf{e}}_{\mathrm{L}}$ is the direction of the path through a thin slice of the inductor wire (of a differential element of the  
path $\mathcal{C}_{\mathrm{L}}$  along the wire) and $\hat{\mathbf{n}}_{\mathrm{L}}$ ($+\hat{\mathbf{e}}_{\mathrm{L}}$ 
or $-\hat{\mathbf{e}}_{\mathrm{L}}$) is the chosen orientation of the cross-section of that slice.

The rules do {\em not} invoke the ``current direction;'' in this way, they are congruent with the current as flux
(Eq.\,\eqref{eq:gendefI}). In addition, they do {\em not}  involve  the direction of the current {\em density} (the drifting direction 
of charge carriers), which simplifies their application, because in general, the current density direction in a wire or a circuit element 
may vary with time and may be difficult to predict before the circuit analysis is completed. 
%
%
\section{Application: analysis of an RLC series circuit}
\label{sec:application}
The application of the rules is demonstrated by analyzing an RLC series circuit. 
\begin{figure}[htbp]
\vskip -2.4cm
\hspace*{-5.cm}
\includegraphics[width=18.5cm]{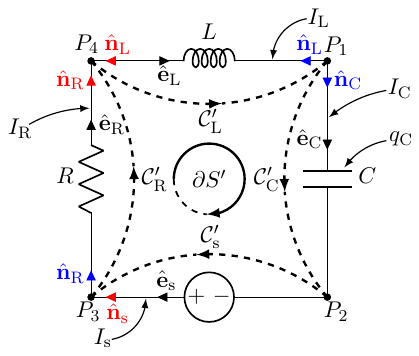}
\vskip -17.cm
\caption{An RLC series circuit with a voltage source. 
The voltage along the path $\mathcal{C}'_{\textnormal{s}}$, passing by the source from $P_2$ to $P_3$, is $U'_{\textnormal{s}}$.
The clockwise orientation, chosen for the loop $\partial S'=\mathcal{C}'_{\textnormal{s}}+\mathcal{C}'_{\textnormal{R}}+
\mathcal{C}'_{\textnormal{L}}+\mathcal{C}'_{\textnormal{C}}$, indicates the  {\em upward} direction $\hat{\mathbf{e}}_{\textnormal{R}}$ 
through the resistor, the {\em rightward} direction $\hat{\mathbf{e}}_{\textnormal{L}}$ along the connecting wires of the inductor, 
the {\em downward} direction $\hat{\mathbf{e}}_{\textnormal{C}}$ through the capacitor, and the {\em leftward} direction 
$\hat{\mathbf{e}}_{\textnormal{s}}$ along the connecting wires of the source. In addition, the {\em leftward} orientations  $
\hat{\mathbf{n}}_{\textnormal{s}}$ and $\hat{\mathbf{n}}_{\textnormal{L}}$ are declared for the wire between the source and 
the junction $P_3$ and for the connecting wires of the inductor,  the {\em upward} orientation $\hat{\mathbf{n}}_{\textnormal{R}}$ 
is set for the resistor connecting wires (and for the resistor terminals),  and the {\em downward} orientation $\hat{\mathbf{n}}_{\textnormal{C}}$ 
is chosen for the wire between the junction $P_1$ and the capacitor. There is no accumulation of charge at any of the four junctions, $\dot{q}^{}_{\textnormal{P}_{\! 1}}
=\ldots =\dot{q}^{}_{\textnormal{P}_{\! 4}}=0$.}
\end{figure}
With all necessary directions set (Fig.\,4), the rules apply as follows: 
%
\begin{description}
%
\item[R.1]{Let $q^{}_{\textnormal{C}}$ be the charge of the {\em upper} capacitor electrode (Fig.\,4), for which $m=1$ (a single wire is 
connected to the electrode) and $k=1$ ($\hat{\mathbf{n}}_{\textnormal{C}}$ points {\em into} the electrode), such that
\begin{equation*}
I_{\textnormal{C}}=+\dot{q}^{}_{\textnormal{C}}.
\end{equation*}

For the junction $P_1$, $m=2$ and $k=0$ (two wires are connected to $P_1$ and their orientations $\hat{\mathbf{n}}_{\textnormal{C}}$ 
and $\hat{\mathbf{n}}_{\textnormal{L}}$ both point {\em away from} $P_1$). Thus, because of $\dot{q}^{}_{\textnormal{P}_{\! 1}}=0$,
$0=-(I_{\textnormal{L}}+I_{\textnormal{C}})$ and
\begin{equation}
\label{eq:apR12}
I_{\textnormal{L}}=-I_{\textnormal{C}}=-\dot{q}^{}_{\textnormal{C}}.
\end{equation}
Similarly, with $\hat{\mathbf{n}}_{\textnormal{L}}$ and $\hat{\mathbf{n}}_{\textnormal{R}}$ both pointing {\em toward} $P_4$
($k=2$), $0=I_{\textnormal{R}}+I_{\textnormal{L}}$, or, together with Eq.\;\eqref{eq:apR12}, 
\begin{equation}
\label{eq:apR14}
I_{\textnormal{R}}=-I_{\textnormal{L}}=+\dot{q}^{}_{\textnormal{C}},
\end{equation}
whereas for  $P_3$, $k=1$ and
\begin{equation}
\label{eq:apR15}
I_{\textnormal{s}}=I_{\textnormal{R}}=+\dot{q}^{}_{\textnormal{C}}.
\end{equation}
}
%
\item[R.3]{For $\hat{\mathbf{n}}_{\textnormal{R}}$ and $\hat{\mathbf{e}}_{\textnormal{R}}$ both pointing upward, 
$\hat{\mathbf{n}}_{\textnormal{R}}\cdot\hat{\mathbf{e}}_{\textnormal{R}}=+1$. Then, according to 
Rule\;\eqref{eq:R3} and Eq.\;\eqref{eq:apR14}, 
\begin{equation}
\label{eq:apR3}
U'_{\textnormal{R}} =-\mathop{I_{\textnormal{R}} R}=-\mathop{\dot{q}^{}_{\textnormal{C}} R}.
\end{equation}
}
%
\item[R.4]{As the upper capacitor electrode is connected to the starting point $P_1$ of the path $\mathcal{C}'_{\textnormal{C}}$, 
$q^{}_{\textnormal{C}}=q^{}_{\textnormal{C},1}$ and
\begin{equation}
\label{eq:apR2}
U'_{\textnormal{C}}=-\frac{q^{}_{\textnormal{C}}}{C}.
\end{equation}
}
%
\item[R.5]{For $\mathop{\hat{\mathbf{n}}_{\textnormal{L}}}$ pointing {\em leftward} and $\mathop{\hat{\mathbf{e}}_{\textnormal{L}}}$ pointing {\em rightward},
$\mathop{\hat{\mathbf{n}}_{\textnormal{L}}}\cdot\mathop{\hat{\mathbf{e}}_{\textnormal{L}}}= -1$. Rule\;\eqref{eq:R4} and Eq.\;\eqref{eq:apR12} thus yield
\begin{equation}
\label{eq:apR4}
U'_{\textnormal{L}} =
+L\dot{I}_{\textnormal{L}}=-\ddot{q}^{}_{\textnormal{C}}L .
\end{equation}
}
%
\item[R.2]{For the loop $\partial S'=\mathcal{C}'_{\textnormal{s}}+\mathcal{C}'_{\textnormal{R}}+\mathcal{C}'_{\textnormal{L}}+
\mathcal{C}'_{\textnormal{C}}$, the rule reads $U'_{\textnormal{s}}+U'_{\textnormal{R}}+U'_{\textnormal{L}}+U'_{\textnormal{C}}=0$,
which, together with Eqs.\;\eqref{eq:apR3}--\eqref{eq:apR4}, leads to
\begin{equation}
\label{eq:qCfinalEq}
\ddot{q}^{}_{\textnormal{C}}L +\mathop{\dot{q}^{}_{\textnormal{C}} R}+q^{}_{\textnormal{C}}\frac{1}{C}=U'_{\textnormal{s}}.
\end{equation}
}
%
\end{description}

Equation\,\eqref{eq:qCfinalEq} (of a driven oscillator) does {\em not} depend on the chosen orientations
of the loop $\partial S'$ and of the wires (their cross-sections). With the specified ``initial'' conditions 
(e.g., $q^{}_{\textnormal{C}}(t=0)$ and $\dot{q}^{}_{\textnormal{C}}(t=0)$) and the source voltage $U'_{\textnormal{s}}(t)$,
the solution $q^{}_{\textnormal{C}}(t)$ of the equation is unique. Given $q^{}_{\textnormal{C}}(t)$, the voltages 
$U'_{\textnormal{C}}$, $U'_{\textnormal{R}}$, and $U'_{\textnormal{L}}$ follow from Eqs.\,\eqref{eq:apR3}--\eqref{eq:apR4}, 
and the currents $I_{\textnormal{C}}$, $I_{\textnormal{L}}$, $I_{\textnormal{R}}$, and $I_{\textnormal{s}}$  follow from
Eqs.\,\eqref{eq:apR12} and \eqref{eq:apR15}, which completes the analysis.

\appendix*
%
%
\section*{Appendix: Derivation of the rules}
\label{app:derivations}
\setcounter{equation}{0}
%
%
\subsection{Derivation of Rule (R.1)}
\label{app:derR1}
Let $n$ be the number per volume (number density) of particles, each having charge $q_1$, drifting with the average velocity 
$\mathbf{v}$. The current density $\mathbf{j}$ is\cite{feynmanj,purcellj}
\begin{equation}
\label{eq:defj}
\mathbf{j}\coloneqq q_1\hskip -0.30mm n \mathbf{v} = q_1\hskip -0.3mm n v\mathop{\hat{\mathbf{e}}_{\mathbf{v}}},
\end{equation}
where $\hat{\mathbf{e}}_{\mathbf{v}}=\mathbf{v}/v$ is the direction of $\mathbf{v}$ and $v$ is its magnitude. 
%
%
%

For simplicity, suppose that all charges drift with the same velocity and consider the charges in a prism 
with the lateral edges parallel to $\mathbf{v}$ (Fig.\,5). The length of the edges is $v\mathop{\od t}$ (
the distance each of the charge carriers will travel in time ${\od t}$) and $A$ is the area of each of the 
two base faces. Let $S$ be the base face {\em toward} which the charges are drifting and let 
$\mathbf{S}=A\hat{\mathbf{n}}$ be the surface vector whose orientation $\hat{\mathbf{n}}$ is 
either $\hat{\mathbf{n}}_{\textnormal{f}}$ (``forward'' with respect to $\mathbf{v}$, such that 
$\hat{\mathbf{n}}_{\textnormal{f}}\cdot \hat{\mathbf{e}}_{\mathbf{v}}> 0$), or $\hat{\mathbf{n}}_{\textnormal{b}}
=-\hat{\mathbf{n}}_{\textnormal{f}}$ (``backward'').
\begin{figure}[htbp]
\vskip -2.cm
\hspace*{-4.0cm}
\includegraphics[width=17cm]{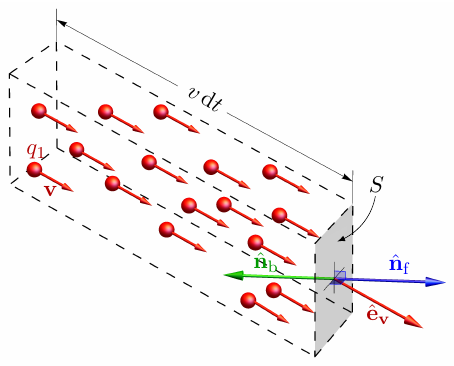}
\vskip -15.45cm 
\caption{Charge carriers in a prism, drifting with velocity $\mathbf{v}$ toward the prism base face $S$; the velocity
is parallel to the lateral edges of the prism.}
\end{figure}
The volume of the prism is
\begin{equation*}
\label{eq:DV}
\od V'\! =
\begin{cases}
+A\mathop{v\mathop{\od t}}\mathop{\hat{\mathbf{n}}\cdot\hat{\mathbf{e}}_{\mathbf{v}}}; & \hskip -2mm \hat{\mathbf{n}}=\hat{\mathbf{n}}_{\textnormal{f}} \\
-A\mathop{v\mathop{\od t}}\mathop{\hat{\mathbf{n}}\cdot\hat{\mathbf{e}}_{\mathbf{v}}}; & \hskip -2mm \hat{\mathbf{n}}=\hat{\mathbf{n}}_{\textnormal{b}} 
\end{cases}\!
=
\begin{cases}
+\mathop{\mathbf{v}\cdot\mathbf{S}}\mathop{\od t} ; & \hskip -2mm \hat{\mathbf{n}}=\hat{\mathbf{n}}_{\textnormal{f}} \\
- \mathop{\mathbf{v}\cdot\mathbf{S}}\mathop{\od t}; & \hskip -2mm \hat{\mathbf{n}}=\hat{\mathbf{n}}_{\textnormal{b}}
\end{cases}
\! ,
\end{equation*}
the number $\od N$ of drifting charges in the prism is $n\mathop{\od V'}$, and their total charge is 
\begin{equation}
\label{eq:Dq}
\od q = q_1\mathop{\od N} =
\begin{cases}
+{I}\mathop{\od t} ; & \hskip -2mm \hat{\mathbf{n}}=\hat{\mathbf{n}}_{\textnormal{f}} \\
- {I}\mathop{\od t}; & \hskip -2mm \hat{\mathbf{n}}=\hat{\mathbf{n}}_{\textnormal{b}}
\end{cases},
\end{equation}
which coincides with the quantity of charge that passes through the surface $S$ in time ${\od t}$.

When $S$ is part of a closed surface $\partial V$ with the interior $V$, the charge, drifting through $S$, causes 
a charge change ${\od q^{}_{\partial V}}$ in $V$. The charge drifts either into $V$ or out of it: 
\begin{enumerate}[label=\alph*)]
%
\item When the charge dries {\em into} $V$ (when the prism is on the {\em outer} side of $V$), the resulting charge change 
${\od q^{}_{\partial V}}$ in $V$ coincides with the charge $\od q$, passed through the surface $S$  (${\od q_{\partial V}}=\od q$),
$\hat{\mathbf{n}}_{\textnormal{f}}$ points {\em into} $V$, and  $\hat{\mathbf{n}}_{\textnormal{b}}$ points {\em out of } it.
%
\item When the charge dries {\em out of} $V$ (when the prism is {\em inside} of $V$), ${\od q^{}_{\partial V}}=-\od q$,
$\hat{\mathbf{n}}_{\textnormal{f}}$ points {\em out of} $V$, and  $\hat{\mathbf{n}}_{\textnormal{b}}$ points {\em into} it.
\end{enumerate}
%
Then, according to Eq.\;\eqref{eq:Dq}, in the {\em both} cases, regardless of whether the charges are drifting {\em into} $V$ 
or {\em out of} it,
\begin{equation}
\label{eq:Dq'S'}
{\od q^{}_{\partial V}}=
\begin{cases}
+I\mathop{\od t} ; & \hskip -1mm \hat{\mathbf{n}}\textnormal{ into }V \\
- {I}\mathop{\od t} ; & \hskip -1mm \hat{\mathbf{n}}\textnormal{ out of }V
\end{cases}.
\end{equation}

Suppose now that  $V$ is a junction of $m$ wires, with $I_i$ being the electric current through the $i$-th wire (Fig.\,1);
the wires (their contact surfaces with $V$) numbered $1$ to $k$ are oriented toward $V$, and the remaining wires 
(contact surfaces) are oriented outward. The charge change $\mathop{\od q^{}_{\partial V}}$ in $V$ due to the charges, drifting 
through $\partial V$, is the sum of the contributions
from particular currents; according to Eq.\,\eqref{eq:Dq'S'} and to the chosen wire orientation, this is
\begin{equation}
\label{eq:Dq'dV'}
\od q^{}_{\partial V} =\left( \sum_{i=1}^{k}I_i -\sum_{i=k+1}^{m}\!I_i\right)\mathop{\od t} .
\end{equation}
Rule\,\eqref{eq:R1} follows from Eq.\,\eqref{eq:Dq'dV'} and from the charge conservation law, according to which the charge 
change $\mathop{\od q^{}_{\partial V}}$ in $V$ due to the charge, passed through its surface, coincides with the {\em total} 
charge change $\od q^{}_{V}$ in $V$. 
%
%
\subsection{Derivation of Rule (R.2)}
\label{app:derR2}
Consider an electric circuit consisting of lumped elements (Fig.\,2), and a loop $\partial S'$ consisting of the paths $\mathcal{C}'_i$
{\em by} the elements, $\partial S' = \sum_i \mathcal{C}'_{i}$. The loop is the boundary of a surface $S'$ with zero time-varying 
magnetic flux, $\dot{\Phi}_{\mathbf{B}}(S')=0$.

The voltage along the path $\mathcal{C}'_i$ is $U'_i$, and the voltage around the loop is
\begin{equation}
\label{eq:loopvoltage}
U(\partial S')=-\!\!\!\oint\limits_{\partial S_{\mathrm{C}}}\!\mathbf{E}\cdot\mathop{\od\mathbf{s}}
                   =\sum\limits_i \Bigl( -\! \int\limits_{\mathcal{C}'_i} \!\mathbf{E}\cdot\mathop{\od\mathbf{s}}\Bigr)
                   =\sum\limits_i U'_i.
\end{equation}
On the other hand, according to the Stokes theorem,
\begin{equation*}
\!\oint\limits_{\partial S'}\!\mathbf{E}\cdot\mathop{\od\mathbf{s}}=
\!\begin{cases}
\!+\!\int_{S'}\!\left(\nabla\times\mathbf{E}\right)\!\cdot\mathop{\od\mathbf{S}};&\hskip -2.7mm \textnormal{positively oriented } \partial S'\\
\!-\!\int_{S'}\!\left(\nabla\times\mathbf{E}\right)\!\cdot\mathop{\od\mathbf{S}};&\hskip -2.7mm \textnormal{negatively oriented } \partial S'
\end{cases}\!,
\end{equation*}
and by the Maxwell--Faraday equation,
\begin{equation*}
\int\limits_{S'}\!\!\left(\nabla\times\mathbf{E}\right)\cdot\mathop{\od\mathbf{S}}=
-\!\!\int\limits_{S'}\!\!\frac{\partial\mathbf{B}}{\partial t}\cdot\mathop{\od\mathbf{S}}
=-\!\mathop{\frac{\od}{\od t}}\!\int\limits_{S'}\!\mathbf{B}\cdot\mathop{\od\mathbf{S}}
=\!-\dot{\Phi}_{\mathbf{B}}(S');
\end{equation*}
above,  integration and differentiation can be interchanged because the region of integration $S'$ does {\em not} vary with time.
Hence,
\begin{equation}
\label{eq:loop}
U(\partial S')
=\!
\begin{cases}
\! +\dot{\Phi}_{\mathbf{B}}(S');&\hskip -2.5mm \textnormal{positively oriented\;} \partial S'\\
\! -\dot{\Phi}_{\mathbf{B}}(S');&\hskip -2.5mm \textnormal{negatively oriented\;} \partial S'
\end{cases},
\end{equation}
which, together with Eq.\,\eqref{eq:loopvoltage} and because of the vanishing $\dot{\Phi}_{\mathbf{B}}(S')$, leads to Rule\,\eqref{eq:R2}.
%
%
\subsection{Derivation of Rule (R.3)}
\label{app:derR3}
Consider a resistor of the form of a homogeneous right cylinder (Fig.\,6), made of isotropic material of resistivity $\zeta$, 
of length $\ell$, with ideally conducting terminals (cylinder base faces) of area $A_{\mathrm{R}}$ each,
and with ideally conducting connecting wires, and suppose that current is driven through the resistor by a uniform axial electric field $\mathbf{E}$; 
throughout the paper,  $\zeta$ is assumed independent of the field magnitude.
\begin{figure}[htbp]
\includegraphics[width=7cm]{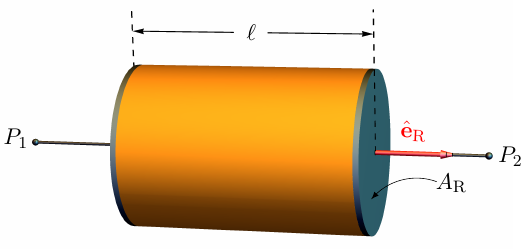}
\vskip -0.2cm
\caption{A resistor of the form of a homogeneous right cylinder. 
The orientation $\hat{\mathbf{e}}_{\mathrm{R}}$ of the path $\mathcal{C}_{\mathrm{R}}$ (straight line) from point $P_1$, 
along the connecting wires and through the resistor to $P_2$, is collinear with the orientation $\hat{\mathbf{n}}_{\mathrm{R}}$ 
of the wires (their cross-sections) and of the resistor terminals, and with the direction $\hat{\mathbf{e}}_{\mathbf{j}}$ of the current 
density through the resistor and the wires.} 
\end{figure}
The voltage $U_{\mathrm{R}}$ along the path $\mathcal{C}_{\mathrm{R}}$ from $P_1$, along the wires and {\em through} the resistor 
to $P_2$, is
\begin{equation}
\label{eq:voltageexam}
U_{\mathrm{R}}= -\mathop{\ell}\mathop{\mathbf{E}\cdot\hat{\mathbf{e}}_{\mathrm{R}}},
\end{equation}
where $\hat{\mathbf{e}}_{\mathrm{R}}$ points from $P_1$ to $P_2$.

In isotropic materials,  the current density direction coincides with that of the field,
\begin{equation}
\label{eq:micohm}
\mathbf{E}=\zeta\,\mathbf{j}.
\end{equation}
Accordingly, a uniform axial $\mathbf{E}$  implies a uniform axial $\mathbf{j}$, signifying that Eq.\;\eqref{eq:currexam} applies for the 
electric current $I_{\mathrm{R}}$ through the resistor (through a flat base of the cylinder). In addition, for the collinear current density 
direction $\hat{\mathbf{e}}_{\mathbf{j}}$ and orientation $\hat{\mathbf{n}}_{\mathrm{R}}$ of the resistor terminals, 
$(\hat{\mathbf{e}}_{\mathbf{j} }\cdot \hat{\mathbf{n}}_{\mathrm{R}})(\hat{\mathbf{e}}_{\mathbf{j} }\cdot \hat{\mathbf{n}}_{\mathrm{R}})=+1$;
thus, multiplying both sides of Eq.\,\eqref{eq:currexam} by $\hat{\mathbf{e}}_{\mathbf{j}}\cdot\hat{\mathbf{n}}_{\mathrm{R}}$ yields
\begin{equation}
\label{eq:jodI}
 jA_{\mathrm{R}}=\lvert I_{\mathrm{R}}\rvert = I_{\mathrm{R}}\mathop{\hat{\mathbf{e}}_{\mathbf{j} }\cdot \hat{\mathbf{n}}_{\mathrm{R}}}.
\end{equation}
From Eqs.\,\eqref{eq:voltageexam}--\eqref{eq:jodI} it then follows that
\begin{equation}
\label{eq:UR}
U_{\mathrm{R}} 
       =-\frac{\ell \zeta}{A_{\mathrm{R}}} I_{\mathrm{R}} 
          \mathop{(\hat{\mathbf{e}}_{\mathbf{j} }\cdot \hat{\mathbf{n}}_{\mathrm{R}})(\hat{\mathbf{e}}_{\mathbf{j}}\cdot\hat{\mathbf{e}}_{\mathrm{R}})}.
\end{equation}
For the collinear $\hat{\mathbf{e}}_{\mathrm{R}}$ and $\hat{\mathbf{n}}_{\mathrm{R}}$,
$(\hat{\mathbf{n}}_{\mathrm{R}}\cdot\hat{\mathbf{e}}_{\mathrm{R}}) (\hat{\mathbf{n}}_{\mathrm{R}}\cdot\hat{\mathbf{e}}_{\mathrm{R}})=+1$,
and thus
\begin{equation*}
(\hat{\mathbf{e}}_{\mathbf{j} }\cdot \hat{\mathbf{n}}_{\mathrm{R}})(\hat{\mathbf{e}}_{\mathbf{j}}\cdot\hat{\mathbf{e}}_{\mathrm{R}})
=
(\hat{\mathbf{e}}_{\mathbf{j} }\cdot \hat{\mathbf{n}}_{\mathrm{R}})(\hat{\mathbf{e}}_{\mathbf{j}}\cdot\hat{\mathbf{e}}_{\mathrm{R}})
(\hat{\mathbf{n}}_{\mathrm{R}}\cdot\hat{\mathbf{e}}_{\mathrm{R}})(\hat{\mathbf{n}}_{\mathrm{R}}\cdot\hat{\mathbf{e}}_{\mathrm{R}}) .
\end{equation*}
In addition, the three collinear vectors, $\hat{\mathbf{e}}_{\mathbf{j} }$, $\hat{\mathbf{n}}_{\mathrm{R}}$, and $\hat{\mathbf{e}}_{\mathrm{R}}$,
occurring in the product 
\begin{equation}
\label{eq:B2}
(\hat{\mathbf{e}}_{\mathbf{j} }\cdot \hat{\mathbf{n}}_{\mathrm{R}})(\hat{\mathbf{e}}_{\mathbf{j}}\cdot\hat{\mathbf{e}}_{\mathrm{R}})
(\hat{\mathbf{n}}_{\mathrm{R}}\cdot\hat{\mathbf{e}}_{\mathrm{R}}),
\end{equation}
either all point in the same direction, or two of them point in the same and one in the opposite direction.
In the former case, all three factors in \eqref{eq:B2} are $+1$, while in the latter case one of the factors is $+1$ and two of 
them are $-1$, meaning that in the both cases the product \eqref{eq:B2} is $+1$.
Cumulatively, this signifies that
\begin{equation}
\label{eq:finally}
(\hat{\mathbf{e}}_{\mathbf{j} }\cdot \hat{\mathbf{n}}_{\mathrm{R}})(\hat{\mathbf{e}}_{\mathbf{j}}\cdot\hat{\mathbf{e}}_{\mathrm{R}})=
\hat{\mathbf{n}}_{\mathrm{R}}\cdot\hat{\mathbf{e}}_{\mathrm{R}},
\end{equation}
and Eq.\,\eqref{eq:UR} simplifies to
\begin{equation}
\label{eq:macohm}
U_{\mathrm{R}}\ =-\hat{\mathbf{n}}_{\mathrm{R}}\cdot\hat{\mathbf{e}}_{\mathrm{R}}\mathop{I_{\mathrm{R}} R},
\end{equation}
where $R={\ell\zeta}/{A_{\mathrm{R}}}$ is the resistance of the element. (For a resistor of different shape and composition, 
the explicit expression for the resistance would be different, but Eq.\,\eqref{eq:macohm} would retain its form.)

Consider now a loop $\partial S_{\mathrm{R}} = \mathcal{C}_{\mathrm{R}} +(-\mathcal{C}'_{\mathrm{R}})$ (Fig.\,3.(a)) 
consisting of the path $\mathcal{C}_{\mathrm{R}}$  and a path $-\mathcal{C}'_{\mathrm{R}}$ {\em by} the resistor,  
from $P_2$ back to $P_1$; the loop is the boundary of a surface $S_{\mathrm{R}}$ with negligible time-varying magnetic 
flux through it. The voltage around the loop is $U(\partial S_{\mathrm{R}})=U_{\mathrm{R}} - U'_{\mathrm{R}},$
where $U'_{\mathrm{R}}=U(\mathcal{C}'_{\mathrm{R}})=-U(-\mathcal{C}'_{\mathrm{R}})$.
At the same time, as for $U(\partial S')$ in Eq.\,\eqref{eq:loop}, above, $U(\partial S_{\mathrm{R}})=\pm 
\dot{\Phi}_{\mathbf{B}}(S_{\mathrm{R}})$; thus, because of the vanishing $\dot{\Phi}_{\mathbf{B}}(S_{\mathrm{R}})$,
$U(\partial S_{\mathrm{R}})=0$ and $U'_{\mathrm{R}} = U_{\mathrm{R}}$, which, together 
with Eq.\,\eqref{eq:macohm}, yields Rule\,\eqref{eq:R3}.
%
\subsection{Derivation of Rule (R.4)}
\label{app:derR4}
Within the quasi-static approximation, the (uniform) electric field between the oppositely charged plates of an ideal parallel-plate capacitor is
\begin{equation*}
\label{eq:capacitorfield}
\mathbf{E}=\frac{q_{\mathrm{C},1}}{\varepsilon A_{\mathrm{C}}}\hat{\mathbf{n}}_{1\to 2},
\end{equation*}
where $A_{\mathrm{C}}$ is the area of (each of the two sides of) a plate, $\varepsilon$ is the permittivity of the material between the plates,
and $\hat{\mathbf{n}}_{1\to 2}$ is the unit vector, perpendicular to the two plates, pointing from the plate with charge $q_{\mathrm{C},1}$ to 
the plate with charge $q_{\mathrm{C},2}$. Hence, the voltage $U_{\mathrm{C}}=U(\mathcal{C}_{\mathrm{C}})$ along the path 
$\mathcal{C}_{\mathrm{C}}$ from point $P_1$, connected to the plate with charge $q_{\mathrm{C},1}$ (Fig.\,3.(b)), along the connecting 
wires and {\em through} the capacitor to point $P_2$, connected to the plate with charge $q_{\mathrm{C},\,2}(=\!-q_{\mathrm{C},1})$, is
\begin{equation}
\label{eq:UC}
U_{\mathrm{C}}=-\frac{q_{\mathrm{C},1}}{C}=+\frac{q_{\mathrm{C},2}}{C}\, ,
\end{equation}
where $C=\varepsilon A_{\mathrm{C}}/\ell$ capacitance of the capacitor and $\ell$ is the distance between its plates. (For a real capacitor, 
the explicit expression for its capacitance would change, but the form of Eq.\;\eqref{eq:UC} would remain the same.)
Just like for the voltages across a resistor,  the voltage $U_{\mathrm{C}}$ along the path $\mathcal{C}_{\mathrm{C}}$ 
{\em through} the capacitor coincides with the voltage $U'_{\mathrm{C}}$ along a path $\mathcal{C}'_{\mathrm{C}}$
{\em by} the capacitor (Fig.\,3.(b)), which, together with Eq.\,\eqref{eq:UC}, yields Rule\,\eqref{eq:R4}. 
%
%
\subsection{Derivation of Rule (R.5)}
\label{app:derR5}
Consider an ideal (long, tightly-wound) solenoid (Fig.\,7) with
perfectly conducting wire.
\begin{figure}[htbp]
\label{fig:solenoid}
\vskip -3.25cm
\hspace*{-3.75cm}
\includegraphics[width=16.5cm]{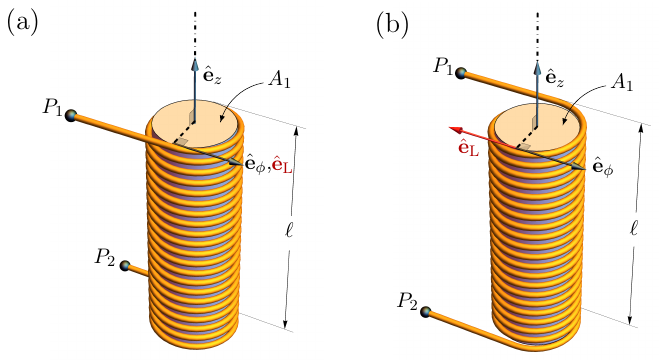}
\vskip -13.75cm
\caption{Two solenoids of length $\ell$ and with $N_{\mathrm{L}}$ turns, each. A single turn $\partial S_1$ defines a flat surface $S_1$
of area $A_1$, coinciding with the cross-sectional area of the solenoids.  Left: the path $\mathcal{C}_{\mathrm{L}}$  along the solenoid wire 
from terminal $P_1$ to terminal $P_2$ is oriented {\em anticlockwise}, looking from above the solenoid. Right:
the path is oriented {\em clockwise.}
}
\end{figure}
By Eq.\;\eqref{eq:micohm}, there is no electric field inside such a wire; 
hence, the voltage $U_{\mathrm{L}}$ (minus the line integral of the {\em zero} field) along the path 
$\mathcal{C}_{\mathrm{L}}$  from $P_1$, along the wire (always staying inside the wire), 
to $P_2$, is zero.\!\cite{feynmanlumped}
%
%

Due to the {\em non-vanishing} time-varying magnetic flux through the solenoid, however, this is {\em not}, in general, equal
to the voltage $U'_{\mathrm{L}}$ along a path $\mathcal{C}'_{\mathrm{L}}$ {\em by} the solenoid.
Consider a loop 
\begin{equation}
\label{eq:partialSL}
\partial S_{\mathrm{L}} = \mathcal{C}_{\textnormal{L}}+(-\mathcal{C}'_{\textnormal{L}}),
\end{equation}
the boundary of a surface $S_{\mathrm{L}}$ (Fig.\,3.(c)).
Within the lumped-element approximation,  
the total magnetic flux $\Phi_{\mathbf{B}}(S_{\mathrm{L}})$ through $S_{\mathrm{L}}$ coincides with the flux  
through
the solenoid, which in turn is the flux $\Phi_{\mathbf{B}}(S_1)$ through the flat surface $S_1$, defined by a single turn (loop $\partial S_1$ of the wire), times the number
$N_{\mathrm{L}}$ of such surfaces (turns),
\begin{equation}
\label{eq:PhiBL}
\Phi_{\mathbf{B}}(S_{\mathrm{L}})
= N_{\mathrm{L}}\Phi_{\mathbf{B}}(S_1).
\end{equation}
For the uniform solenoid field 
%
$\mathbf{B}=B\mathop{\hat{\mathbf{e}}_{\mathbf{B}}}$
%
and the flat surface $S_{1}$, the flux $\Phi_{\mathbf{B}}(S_{1})$ is
\begin{equation*}
\Phi_{\mathbf{B}}(S_{1})=\mathbf{B}\cdot\mathbf{S}_1,
\end{equation*}
where $\mathbf{S}_1=A_1\hat{\mathbf{n}}_1$ is the surface vector with magnitude (area of $S_{1}$) $A_1$ 
and direction (orientation of $S_{1}$) $\hat{\mathbf{n}}_1$,
\begin{equation}
\label{eq:Bmag}
B=\frac{\mu N_{\mathrm{L}} \lvert I_{\mathrm{L}}\rvert }{\ell}
\end{equation}
is the (quasi-static approximation of the) field magnitude, 
$\lvert I_{\mathrm{L}}\rvert$ is the current {\em magnitude} ({\em not} current) through the solenoid wire, 
and $\mu$ is the permeability of the solenoid core;  throughout the present analysis, $\mu$ is a positive scalar constant. 
(The field magnitude\;\eqref{eq:Bmag} is deducible from the Biot-Savart law and is further discussed in the Supplementary Materials.)

The orientation $\hat{\mathbf{n}}_1$ of $S_1$ and the direction $\hat{\mathbf{e}}_{\mathbf{B}}$ of $\mathbf{B}$ are both
collinear with the solenoid axis, which in turn
coincides with the longitudinal axis of a cylindrical coordinate system
with the basis vector $\hat{\mathbf{e}}_z$ pointing upward (Fig.\,7). Then,  $\hat{\mathbf{n}}_1$ can be {\em chosen}  either $+\hat{\mathbf{e}}_z$ or
$-\hat{\mathbf{e}}_z$, whereas $\hat{\mathbf{e}}_{\mathbf{B}}$ is determined by the current {\em density} direction ({\em not} current direction) 
along the solenoid wire:  it points upward if 
$\hat{\mathbf{e}}_{\mathbf{j}}=\hat{\mathbf{e}}_{\phi}$ and points downward if $\hat{\mathbf{e}}_{\mathbf{j}}=-\hat{\mathbf{e}}_{\phi}$,
\begin{equation*}
\hat{\mathbf{e}}_{\mathbf{B}}=(\hat{\mathbf{e}}_{\mathbf{j}}\cdot\hat{\mathbf{e}}_{\phi})\mathop{\hat{\mathbf{e}}_z}.
\end{equation*}

As in Eq.\;\eqref{eq:jodI}, because  $\hat{\mathbf{e}}_{\mathbf{j}}$ in a thin slice of the solenoid wire and the orientation
$\hat{\mathbf{n}}_{\mathrm{L}}$ of the slice cross-section are collinear,
\begin{equation*}
\label{eq:absI}
\lvert I_{\mathrm{L}}\rvert =I_{\mathrm{L}}\mathop{\hat{\mathbf{e}}_{\mathbf{j}}\cdot\hat{\mathbf{n}}_{\mathrm{L}}},
\end{equation*}
and the flux $\Phi_{\mathbf{B}}(S_1)$ reads
\begin{equation*}
\Phi_{\mathbf{B}}(S_1)=\frac{\mu N_{\mathrm{L}} A_1}{\ell} I_{\mathrm{L}}(\mathop{\hat{\mathbf{n}}_{\mathrm{L}} \cdot\hat{\mathbf{e}}_{\mathbf{j}}})(\mathop{\hat{\mathbf{e}}_{\mathbf{j}} \cdot\hat{\mathbf{e}}_{\phi}}){(\hat{\mathbf{e}}_{z} \cdot\hat{\mathbf{n}}_{1})}.
\end{equation*}
This, because of 
\begin{equation}
\label{eq:finally1}
(\mathop{\hat{\mathbf{n}}_{\mathrm{L}} \cdot\hat{\mathbf{e}}_{\mathbf{j}}})(\mathop{\hat{\mathbf{e}}_{\mathbf{j}} \cdot\hat{\mathbf{e}}_{\phi}})=
\mathop{\hat{\mathbf{n}}_{\mathrm{L}} \cdot\hat{\mathbf{e}}_{\phi}},
\end{equation}
simplifies to
\begin{equation}
\label{eq:fluxcoil1}
\Phi_{\mathbf{B}}(S_1)
=\frac{L\ }{N_{\mathrm{L}}} I_{\mathrm{L}} \mathop{(\hat{\mathbf{n}}_{\mathrm{L}} \cdot\hat{\mathbf{e}}_{\phi})}
                                                                    {(\hat{\mathbf{e}}_{z} \cdot\hat{\mathbf{n}}_{1})} ,
\end{equation}
where $L= \mu N_{\mathrm{L}}^2 A_1/\ell$ is the inductance of the solenoid. 
(The proof of Eq.\;\eqref{eq:finally1} is analogous to that of Eq.\;\eqref{eq:finally}, above.)

By analogy with Eq.\,\eqref{eq:loop} and according to Eq.\,\eqref{eq:PhiBL},
\begin{equation}
\label{eq:loop1}
U(\partial S_{\mathrm{L}})=
\begin{cases}
+N_{\mathrm{L}}\dot{\Phi}_{\mathbf{B}}(S_1);&\hskip -2.5mm \textnormal{pos.\;orient.\;} \partial S_1\\
-N_{\mathrm{L}}\dot{\Phi}_{\mathbf{B}}(S_1);&\hskip -2.5mm \textnormal{neg.\;orient.\;} \partial S_1
\end{cases}\!.
\end{equation}
Whether a loop $\partial S_1$ is positively or negatively oriented depends on the orientations $\hat{\mathbf{n}}_1$ 
of $\mathbf{S}_1$ and $\hat{\mathbf{e}}_{\mathrm{L}}$ of (a differential element of) the path $\mathcal{C}_{\mathrm{L}}$ along the solenoid wire:
\begin{enumerate}[label=\alph*)]
\item When $\hat{\mathbf{e}}_{\mathrm{L}}=\hat{\mathbf{e}}_{\phi}$ (Fig.\,7.(a)), $\partial S_1$ is {\em positively} oriented if 
$\hat{\mathbf{n}}_1=+\hat{\mathbf{e}}_z$, and
is {\em negatively} oriented if $\hat{\mathbf{n}}_1=-\hat{\mathbf{e}}_z$. 
\item When $\hat{\mathbf{e}}_{\mathrm{L}}=-\hat{\mathbf{e}}_{\phi}$ (Fig.\,7.(b)), $\partial S_1$ is {\em positively} oriented
if $\hat{\mathbf{n}}_1=-\hat{\mathbf{e}}_z$, and is {\em negatively} oriented if $\hat{\mathbf{n}}_1=+\hat{\mathbf{e}}_z$. 
\end{enumerate}
From Eq.\;\eqref{eq:fluxcoil1} it then follows that
\begin{equation}
\label{eq:PhiB1cases}
\Phi_{\mathbf{B}}(S_1) =
\begin{cases}
+\frac{L\ }{N_{\mathrm{L}}} I_{\mathrm{L}}\mathop{ \hat{\mathbf{n}}_{\mathrm{L}}\! \cdot\hat{\mathbf{e}}_{\mathrm{L}}} ; 
&\hskip -2.75mm \textnormal{pos.\;orient.\;} \partial S_1\\
-\frac{L\ }{N_{\mathrm{L}}} I_{\mathrm{L}} \mathop{\hat{\mathbf{n}}_{\mathrm{L}} \!\cdot\hat{\mathbf{e}}_{\mathrm{L}}}; &
\hskip -2.75mm \textnormal{neg.\;orient.\;} \partial S_1
\end{cases}\!\!,
\end{equation}
which, when combined with Eq.\;\eqref{eq:loop1}, yields
\begin{equation}
\label{eq:loopfinal}
U(\partial S_{\mathrm{L}})=
\mathop{\hat{\mathbf{n}}_{\mathrm{L}} \cdot\hat{\mathbf{e}}_{\mathrm{L}}}L\dot{I}_{\mathrm{L}}\, . 
\end{equation}

On the other hand, by the decomposition\;\eqref{eq:partialSL} of $\partial S_{\mathrm{L}}$, 
$U(\partial S_{\mathrm{L}})= U_{\mathrm{L}}-U'_{\mathrm{L}}$, which,
because of the vanishing $U_{\mathrm{L}}$ 
and together with Eq.\,\eqref{eq:loopfinal},
gives Rule\,\eqref{eq:R5}.

The relative sign between $U'_{\textnormal{L}}$ and $L\dot{I}_{\textnormal{L}}$ in \eqref{eq:R5} depends only on the {\em relative} orientation $\hat{\mathbf{e}}_{\textnormal{L}}$
of the path through a thin slice of the solenoid wire with respect to the chosen orientation $\hat{\mathbf{n}}_{\textnormal{L}}$ of (the cross-section of)
that slice. While each of the two directions varies from one slice to another, their relative orientation (the product 
$\hat{\mathbf{n}}_{\textnormal{L}} \cdot\hat{\mathbf{e}}_{\textnormal{L}}$) does {\em not}, and coincides with the relative orientation
(the dot product) of $\hat{\mathbf{n}}_{\textnormal{L}}$ and $\hat{\mathbf{e}}_{\textnormal{L}}$ in the connecting wires.

Again, for an inductor of a different shape, the explicit expression for its inductance would differ from that of a solenoid, but the form of 
\eqref{eq:R5} would remain unchanged.
%



\clearpage
\raggedbottom
\begin{center}
\section*{Supplementary materials}
\end{center}
\appendix
\setcounter{section}{19}
\setcounter{equation}{0}
%
%
\subsection*{The Biot-Savart law and the solenoid magnetic field}
\label{app:biotsavart}

The magnetic field $\mathbf{B}(\mathbf{r})$,
generated by a charge flow, is\cite{feynmanB,griffithsB} 
\begin{equation}
\label{eq:biot-savart}
\mathbf{B}(\mathbf{r})=\frac{\mu_0}{4\pi}\int\limits_{\mathbb{R}^3}
\frac{\mathbf{j}(\mathbf{r}')\times(\mathbf{r}-\mathbf{r}')}{{\lvert \mathbf{r}-\mathbf{r}'\rvert}^3} \mathop{\od V'},
\end{equation}
where $\mathbf{j}(\mathbf{r}')$ and $\od V'$ are the current density and the volume element at position $\mathbf{r}'$, and $\mu_0$ is
the magnetic constant. The expression is exact for a steady flow, and is approximately correct for a slowly varying flow, 
in the quasi-static approximation.

Suppose the flow is restricted to a wire, and let $\mathcal{C}$ be a streamline of the flowing charge carriers ($\hskip 0.3mm \mathbf{j}$ is tangential to $\mathcal{C}$ 
at each of its points $\mathbf{r}_{\mathcal{C}}(\ell)$). The wire can be decomposed
into short sections (slices) of length $\od \ell$; at every point, the (flat) base face $S$ of a slice (Fig.\,8) is perpendicular to 
$\mathbf{j}$, or equivalently,  the orientation vector $\hat{\mathbf{n}}$ of $S$ and $\mathbf{j}$ are collinear, 
such that $\hat{\mathbf{e}}_{\mathbf{j}}\cdot\hat{\mathbf{n}}=\pm 1$.
\begin{figure}[hb]
\label{fig:biotsavart}
\vskip -0.15cm
\hspace*{-4.35cm}
\includegraphics[width=17cm]{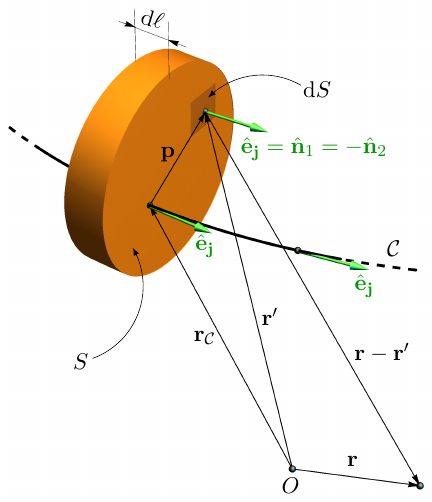}
\vskip -15.5cm 
\caption{A slice of a wire and a streamline $\mathcal{C}$ of the flowing charge carriers. The orientation $\hat{\mathbf{n}}$ of the slice base face $S$ 
is either $\hat{\mathbf{n}}_1$ or $\hat{\mathbf{n}}_2$ and is collinear with the current density at every point of $S$. The position $\mathbf{r}'$ of a point on $S$ expresses 
as a sum of the position vector $\mathbf{r}_{\mathcal{C}}$ of the streamline intersection with $S$,
and a vector $\mathbf{p}$, lying on $S$.}
\end{figure}
The position $\mathbf{r}'$ of a point on $S$ expresses as a sum of the position vector $\mathbf{r}_{\mathcal{C}}$ of the streamline intersection with $S$, 
and a vector $\mathbf{p}$, lying on $S$. The field\;\eqref{eq:biot-savart} can then be rewritten as
\begin{equation*}
\mathbf{B}(\mathbf{r})=\frac{\mu_0}{4\pi}\int\limits_{\mathcal{C}}\left[\ \int\limits_{S}
\frac{\mathbf{j}(\mathbf{r}_{\mathcal{C}}+\mathbf{p})\times(\mathbf{r}-\mathbf{r}_{\mathcal{C}}-\mathbf{p})}{{\lvert \mathbf{r}-\mathbf{r}_{\mathcal{C}}-\mathbf{p}\rvert}^3} \mathop{\od S}\right]\mathop{\od \ell},
\end{equation*}
where $\od S$ is a differential element of $S$.
For a thin wire for which $\lvert\mathbf{p}\rvert \ll \lvert \mathbf{r}-\mathbf{r}_{\mathcal{C}}\rvert$ holds for {\em all} points on the base faces of {\em all} slices,
%
\begin{equation*}
\frac{\mathbf{r}-\mathbf{r}_{\mathcal{C}}-\mathbf{p}}{\hskip 1.6mm{\lvert \mathbf{r}-\mathbf{r}_{\mathcal{C}}-\mathbf{p}\rvert}^3}\approx
\frac{\mathbf{r}-\mathbf{r}_{\mathcal{C}}}{\hskip 1.6mm{\lvert \mathbf{r}-\mathbf{r}_{\mathcal{C}}\rvert}^3},
\end{equation*} 
and hence
\begin{equation*}
\mathbf{B}(\mathbf{r})\approx\frac{\mu_0}{4\pi}\int\limits_{\mathcal{C}}\left[\ \int\limits_{S}
\mathbf{j}\mathop{\od S}\,\right]\times\frac{(\mathbf{r}-\mathbf{r}_{\mathcal{C}})}{{\hskip 1.6mm\lvert \mathbf{r}-\mathbf{r}_{\mathcal{C}}\rvert}^3}\mathop{\od \ell}.
\end{equation*}
In addition, by Definition\;(1.a) of the electric current,
\begin{equation*}
I=
\int\limits_{S}\! j\mathop{\hat{\mathbf{e}}_{\mathbf{j}}\cdot\hat{\mathbf{n}}}\mathop{\od S}=\pm\int\limits_{S}\! j\mathop{\od S},
\end{equation*}
such that
\begin{equation*}
\int\limits_{S}\mathbf{j}\mathop{\od S}=\hat{\mathbf{e}}_{\mathbf{j}}\int\limits_{S}\! j\mathop{\od S}=\hat{\mathbf{e}}_{\mathbf{j}}\mathop{\lvert I\rvert},
\end{equation*}
and Eq.\,\eqref{eq:biot-savart} reduces to
\begin{equation}
\label{eq:biot-savart1}
\mathbf{B}(\mathbf{r})\approx
\frac{\mu_0}{4\pi}\!\int\limits_{\mathcal{C}}\frac{\mathop{\lvert I\rvert}\mathop{\od\bs{\ell}}\times(\mathbf{r}-\mathbf{r}_{\mathcal{C}}(\ell))}
{{\lvert \mathbf{r}-\mathbf{r}_{\mathcal{C}}(\ell)\rvert}^3},
\end{equation}
where $\od \bs{\ell}=\hat{\mathbf{e}}_{\mathbf{j}}\mathop{\od \ell}$. Equation\;\eqref{eq:biot-savart1} (the Biot-Savart law) and the magnitude of the solenoid  field\;\eqref{eq:Bmag}, deducible from it,
both differ from the corresponding expressions in the literature\cite{feynmanbs,grantbs,purcellbs,griffithsbs,walkerbs} in that they involve current {\em magnitudes}  $\lvert I\rvert$ and $\lvert I_{\textnormal{L}}\rvert$, 
instead of the currents
$I$ and $I_{\textnormal{L}}$ themselves.\\[10.cm]
%
\end{document}